\documentclass[twocolumn,prb,superscriptaddress,longbibliography]{revtex4-2}
\pdfoutput=1
\usepackage{graphicx}
\usepackage{subfigure}
\usepackage{color}
\usepackage{amsmath}
\usepackage{enumitem}
\usepackage{amssymb}
\usepackage[normalem]{ulem}
\usepackage{adjustbox}
\usepackage{bm}
\usepackage{braket}
\usepackage{physics}
\usepackage{dcolumn}
\usepackage{hyperref}
\usepackage{siunitx}
\usepackage{pdfpages}
\makeatletter
\patchcmd{\@outputpage@head}{\@ifx{\LS@rot\@undefined}{}{\LS@rot}}{}{}{}
\makeatother

\hypersetup{
	colorlinks = true,
	linkcolor = blue,
	citecolor = blue,
	urlcolor  = blue,
}

\DeclareMathAlphabet{\mathdutchcal}{U}{dutchcal}{m}{n}
\SetMathAlphabet{\mathdutchcal}{bold}{U}{dutchcal}{b}{n}
\DeclareMathAlphabet{\mathdutchbcal}{U}{dutchcal}{b}{n}
\newcommand{\pp}[0]{\mathbf{p}}

\begin{document}

\title{Enhanced spin-current generation in Dirac altermagnets through Klein tunneling}

\author{Tomas T. Osterholt}
\altaffiliation{These authors contributed equally.}
\affiliation{%
Institute for Theoretical Physics, Utrecht University, 3584CC Utrecht, The Netherlands\\
}%

\author{Lumen Eek}
\altaffiliation{These authors contributed equally.}
\affiliation{%
Institute for Theoretical Physics, Utrecht University, 3584CC Utrecht, The Netherlands\\
}%
\thanks{These authors contributed equally to this work.}

\author{Cristiane Morais Smith}
\affiliation{%
Institute for Theoretical Physics, Utrecht University, 3584CC Utrecht, The Netherlands\\
}

\author{Rembert A. Duine}
\affiliation{%
Institute for Theoretical Physics, Utrecht University, 3584CC Utrecht, The Netherlands\\
}
\affiliation{Department of Applied Physics, Eindhoven University of Technology,
P.O. Box 513, 5600 MB Eindhoven, The Netherlands
}%

\date{\today}

\begin{abstract}
Altermagnets have recently emerged as a new platform for spintronics applications, offering spin-split electronic bands despite vanishing net magnetization. Here, we investigate spin-current generation in Dirac altermagnets and identify Klein tunneling as an efficient mechanism for enhancing spin transport. Using a low-energy Dirac model combined with scattering theory, we demonstrate that Klein tunneling in altermagnets is strongly spin-dependent and can be used to effectively control the electronic spin-current polarization by, for instance, adjusting the height, width and orientation of the potential barrier. Finally, we explore how the $\ell$-wave symmetry of the Dirac altermagnet shapes the spin-current polarization and transmission, focusing especially on the $d$- and $g$-wave cases. Particularly promising results are obtained for the $g$-wave Dirac altermagnet, as it is found that the presence of a potential barrier can significantly boost the spin-current polarization, even when the intrinsic polarization due to the spin-split band structure is vanishingly small. For a barrier implemented via electrostatic gating, such a mechanism would in turn allow the spin-current polarization to be switched on and off via a gate voltage.
\end{abstract}

\maketitle

\section{Introduction}

For decades, one of the core goals of the field of spintronics \cite{Zutic2004,Bader2010,Hirohata2020} has been to exploit the electron's spin degree of freedom to realize information processing devices that are more energy-efficient compared to conventional electronics. Central goals of the field include efficient generation, manipulation, and detection of spin-polarized currents and spin waves, as well as control over their dispersion and transport properties. From a microscopic perspective, spin transport and spin dynamics are governed by the underlying electronic and magnonic band structures, which, respectively, determine spin polarization, group velocities, and selection rules for scattering processes. Consequently, a central theme in contemporary spintronics is magnetic band structure engineering, which involves the deliberate design of spin-dependent electronic and magnonic dispersions.

In recent years, several approaches to manipulate magnetic band structures have been explored. These approaches include, but are not limited to, twist engineering \cite{Li2020,Song2021,Xie2021,Liu2025,Osterholt2025}, electrostatic gating \cite{Jiang2018,Wu2023,Hendriks2024}, optical control \cite{Kirilyuk2010,Elliott2016,Dabrowski2022} and the exploitation of magneto-elastic coupling effects \cite{Bazazzadeh2021,To2023,Gillard2024,Go2026}. Following the foundational work by \v{S}mejkal \textit{et al.} \cite{Smejkal2022}, it was demonstrated that magnetic band structures can also be addressed at the level of magnetic space-group symmetries, giving rise to the concept of altermagnetism. One of the main benefits of altermagnets is that they combine key properties of ferromagnets, such as spin-polarized electronic bands, with the vanishing net magnetization of conventional antiferromagnets, which makes them particularly appealing for spintronics applications. 

As a direct consequence of the spin-split band structure, an electronic spin-polarized current intrinsically arises in an altermagnetic system. A natural question to ask then is whether this spin polarization can still be controlled efficiently by external means without destroying the altermagnetic order. One possible approach would be to exploit a spin-dependent tunneling phenomenon that allows significant control of the spin polarization while maintaining sizable spin currents. 

The above mentioned requirements immediately point towards Klein tunneling as a suitable mechanism. Klein tunneling was originally predicted \cite{Klein1929} in relativistic quantum mechanics as a counterintuitive effect in which an electron can transmit perfectly through a step-like potential barrier whose height $V_0$ is at least twice the electron's rest mass energy. This apparent paradox arises because, for a sufficiently strong potential, the barrier region supports antiparticle states, allowing the electron to traverse it via an intermediate antiparticle mode. Unfortunately, detection of Klein tunneling with elementary particles has thus far proven to be experimentally unfeasible, largely due to the enormous electric fields required to achieve a potential drop on the order of the particle’s rest mass energy over a length scale given by its Compton wavelength \cite{Katsnelson2006}. In condensed-matter systems, however, the existence of effectively massless quasiparticles leads to a physically observable analogue of this phenomenon in materials with linear band crossings, the most prominent example being graphene \cite{Katsnelson2006,Beenakker2008,Stander2009,Young2009,Du2018}. Beyond graphene, Klein tunneling has been explored in other fermionic systems such as topological insulator surface states \cite{Lee2019} and Weyl semimetals \cite{OBrien2016}, as well as in bosonic systems \cite{Wagner2010}, including photons \cite{Zhang2022} and magnons \cite{Harms2022,Yuan2023,Bassant2025}. 

\begin{figure}
    \centering
    \includegraphics[width=\linewidth]{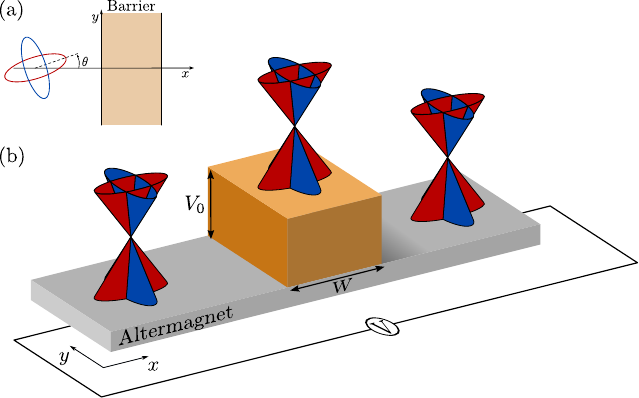}
    \caption{\textbf{Schematic of Klein tunneling in a Dirac altermagnet}. (a) Orientation of the barrier with respect to the altermagnet dispersion. The coordinate system is always chosen in such a way that the barrier is along the $y$-direction. (b) Sketch of the proposed tunnel junction. An $\ell$-wave Dirac altermagnet (here, $\ell = d$) is subjected to a potential barrier of height $V_0$ between $x=0$ and $x=W$, with $W$ denoting the barrier width.  The spin-split electronic bands of the altermagnet are illustrated, with red corresponding to the spin-up (w.r.t. the $z$-axis) band and blue corresponding to the spin-down band. In this case, the angle $\theta$ between the altermagnet and the barrier is chosen to be $0$.
    For $|V_0| \gg |E|$, where $E$ is the electron energy, the system is in the Klein tunneling regime. Applying a voltage difference $V$ across the altermagnet induces a spin-polarized current, whose polarization is strongly affected by Klein tunneling.}
    \label{Figure Setup}
\end{figure}

In this paper, we explore Klein tunneling in Dirac altermagnets, i.e., altermagnets whose electronic band structure hosts symmetry-protected Dirac fermions, and we demonstrate how this effect can be used to control the spin-current polarization. An illustration of the setup is given in Figure \ref{Figure Setup}. We consider a $\ell$-wave Dirac altermagnet, with an applied potential barrier of height $V_0$ present between $x=0$ and $x = W$, where $W$ denotes the barrier width.  We then calculate the zero-bias transmission coefficient, as well as the spin current through the barrier in the presence of an applied voltage difference $V$ between the two sides of the altermagnet. As a main result, we find that the net spin-current polarization can be significantly enhanced due to this potential barrier, while still maintaining a sizable total current. The implications of these findings are that Klein tunneling in Dirac altermagnets can serve as an efficient mechanism to control spin-current polarization, potentially enabling the construction of gate-voltage-controlled spin transport elements.

The outline of the paper is as follows. In Section \ref{Section Minimal Models}, we introduce a minimal model for $\ell$-wave Dirac altermagnets, and we use this model in Section \ref{Section Scattering Theory} to calculate the zero-bias transmission coefficients. In Section \ref{Section Spin Currents}, we then use the Landauer–Büttiker formalism to derive the electronic spin currents and show that, in the Klein tunneling limit, the spin-current polarization depends strongly on the barrier height, width, and orientation, thus enabling tunable control of the spin-current polarization through these parameters. Finally, further research directions and candidate materials are explored in Section \ref{Section Conclusion}.

\section{Minimal models for Dirac altermagnets}\label{Section Minimal Models}
Making use of the framework developed in Ref. \cite{Roig2024}, we begin by introducing minimal models for $\ell$-wave Dirac altermagnets. Here, and throughout the remainder of this work, we choose the spin quantization axis to be the $z$-axis without loss of generality. We start from an ordinary Dirac Hamiltonian with continuous rotational ($C_\infty$) symmetry and we then add terms that break this symmetry. Correspondingly, the Hamiltonian describing the system is given by
\begin{align}
    \hat{\mathcal{H}} &= v_F \, \sigma_0(\mathbf{p} \cdot \boldsymbol{\tau}) + \xi  \begin{pmatrix}
        \mathbf{f}^{(\ell)}_\uparrow(\mathbf{p})\cdot \boldsymbol{\tau} & 0 \\
        0 & \mathbf{f}^{(\ell)}_\downarrow(\mathbf{p})\cdot \boldsymbol{\tau}
    \end{pmatrix} \label{eq:protoDiracH}
    ,
\end{align}
where the first term is a linear isotropic Dirac dispersion with Fermi velocity $v_F$ and the second term represents a spin- and momentum-dependent modulation characterized by the $\ell$-wave form factor $\mathbf{f}_\sigma^{(\ell)}(\pp)$ with strength $\xi$. We have introduced the momentum vector $\mathbf{p} = (\hat{p}_x, \hat{p}_y)$, with $\hat{p}_{x/y}$ the momentum operators corresponding to, respectively, the $x$- and $y$-directions. The spin degree of freedom is indicated by $\sigma$, while the sublattice is denoted by $\tau$. We have also introduced the shorthand notation $\sigma_i \tau_j = \sigma_i \otimes\tau_j$, where $\sigma_j$ ($\tau_j$) with $j \in \{x,y,z\}$ are the Pauli spin-1/2 matrices and where $\sigma_0$ ($\tau_0$) corresponds to the $2 \times 2$ identity matrix.
We emphasize that only the even-parity modes ($\ell = d,g,i$)  correspond to altermagnets \cite{Smejkal2022}. Below, we list one example of spin-dependent form factors $\mathbf{f}_\sigma^{(\ell)}(\mathbf{p})$ that describe the momentum-dependent spin splitting for each mode:
\begin{align}
    \mathbf{f}_\uparrow^{(d)}(\mathbf{p}) &=  p_x \, \hat{\mathbf{x}}-p_y \, \hat{\mathbf{y}}, &
    \mathbf{f}_\downarrow^{(d)}(\mathbf{p}) &=  -\mathbf{f}_\uparrow^{(d)}(\mathbf{p}), \\
    \mathbf{f}_\uparrow^{(g)}(\mathbf{p}) &= (p_x^2 - p_y^2)^2 \, \hat{\mathbf{z}}, &
    \mathbf{f}_\downarrow^{(g)}(\mathbf{p}) &= 4 p_x^2 p_y^2 \, \hat{\mathbf{z}}, \\
    \mathbf{f}_\uparrow^{(i)}(\mathbf{p}) &=  p_x p_y (p_x^2 - p_y^2) \, \hat{\mathbf{z}}, &
    \mathbf{f}_\downarrow^{(i)}(\mathbf{p}) &= - \mathbf{f}_\uparrow^{(i)}(\mathbf{p}).
\end{align}
The $d$-wave Dirac altermagnet is of particular interest, because it can be solved analytically. The corresponding Hamiltonian reads
\begin{align}\label{Equation d-wave Altermagnet Hamiltonian}
    \hat{\mathcal{H}} &= v_F\, \sigma_0(\mathbf{p} \cdot \boldsymbol{\tau}) + \xi \,\sigma_z(\hat{p}_x \tau_x -\hat{p}_y\tau_y) \notag\\
    &= \begin{bmatrix}
        v_x \hat{p}_x \tau_x + v_y \hat{p}_y \tau_y & 0 \\
        0 & v_y \hat{p}_x \tau_x + v_x \hat{p}_y \tau_y \\
    \end{bmatrix}
    ,
\end{align}
with $v_x = v_F +  \xi$ and $v_y = v_F -  \xi$. The spectrum thus consists of two spin-split elliptic Dirac cones that are rotated by $90$ degrees with respect to each other, as illustrated in Fig.~\ref{Figure Setup}.


\section{Transmission Coefficients}\label{Section Scattering Theory}

Here, we derive the spin-dependent transmission coefficients $\mathcal{T}_{\sigma}(\mathbf{p})$ for an $\ell$-wave Dirac altermagnet, with $\ell = d,g$. We first focus on the $d$-wave case, which can be solved analytically. A similar approach can be applied for the $g$-wave case, although numerical methods are then needed to determine $\mathcal{T}_{\sigma}(\mathbf{p})$. Finally, we present plots of the spin-dependent transmission coefficients for both $d$- and $g$-wave Dirac altermagnets, and we demonstrate their strong dependence on parameters such as the barrier height, width and orientation.

\subsection{Analytical calculation for $d$-wave Dirac altermagnets}
First, we analytically solve for the zero-bias transmission coefficient of a Dirac $d$-wave altermagnet using quantum-mechanical scattering theory. As shown in Eq. ~\eqref{Equation d-wave Altermagnet Hamiltonian}, the Hamiltonian of the Dirac $d$-wave altermagnet is diagonal in spin-space, with the terms on the spin diagonal being given by $v_x \hat{p}_x \tau_x + v_y \hat{p}_y \tau_y$ ($\sigma = \uparrow$) and $v_y \hat{p}_x \tau_x + v_x \hat{p}_y \tau_y$ ($\sigma = \downarrow$), respectively. We can therefore focus our attention on the calculation of the transmission coefficient $\mathcal{T}_{\uparrow}(\mathbf{k})$, since $\mathcal{T}_{\downarrow}(\mathbf{k})$ can be obtained directly from said result by making the substitutions $v_x \rightarrow v_y$ and $v_y \rightarrow  v_x$. 

For these reasons, we now tackle the problem of calculating the transmission coefficient $\mathcal{T}_{\uparrow}(\mathbf{k})$ for the effective (sub-)Hamiltonian $\hat{\mathcal{H}} = v_x \hat{p}_x \tau_x + v_y \hat{p}_y \tau_y$. Because of the anisotropy of the energy bands of this (sub-)Hamiltonian, the relative orientation of the potential barrier with respect to the bands is physically relevant, and we therefore also study the more general case of a rotated elliptic Dirac cone. Finally, to keep the notation as concise and clear as possible, we will drop the reference to the spin label $\sigma$ for the remainder of this subsection.

\subsubsection{Elliptic Dirac cones}
The Hamiltonian $\hat{\mathcal{H}}$ giving rise to an elliptic Dirac cone band structure with its center at momentum $\hbar \mathbf{k} = 0$, and its major and minor axes along $k_x$ and $k_y$, is given by
\begin{align}
    \hat{\mathcal{H}} = v_x \hat{p}_x\tau_x + v_y \hat{p}_y \tau_y.
\end{align}
Introducing rotated momentum operators $\hat{p}_{x',y'}$ via
\begin{align} \label{eq:rotp}
    \begin{pmatrix}
        \hat{p}_{x'}\\
        \hat{p}_{y'}
    \end{pmatrix}
    &= 
    \mathcal{R}(\theta)
    \begin{pmatrix}
        \hat{p}_{x}\\
        \hat{p}_{y}
    \end{pmatrix}
    ,
\end{align}
with $\mathcal{R}(\theta)$ the $2\times 2$ rotation matrix,
\begin{align}
    \mathcal{R}(\theta) &=
    \begin{pmatrix}
        \cos \theta & - \sin \theta\\
        \sin \theta & \cos \theta
    \end{pmatrix}
    ,
\end{align}
we can straightforwardly generalize to Hamiltonians corresponding to elliptic Dirac cones rotated by an angle $\theta$ with respect to the $k_x$-axis. Such Hamiltonians $\hat{\mathcal{H}}$ are given by
\begin{align}\label{Equation General Elliptic Dirac Hamiltonian}
    \hat{\mathcal{H}} &= v_x \hat{p}_{x'}\tau_x + v_y \hat{p}_{y'} \tau_y. 
\end{align}
The eigenstates of this Hamiltonian are 2-dimensional Dirac spinors of the form $\boldsymbol{\Psi}(x,y) = e^{i k_x x + i k_y y} \,\boldsymbol{\Psi}_0$, with $\boldsymbol{\Psi}_0$ a two-component vector that does not depend on $x$ and $y$, and their corresponding energies $E$ are given by
\begin{align}\label{Equation Dirac Cone Energy}
    E^2(\mathbf{k}) &= \hbar^2 v_x^2 k_{x'}^2(\theta) + \hbar^2 v_y^2 k_{y'}^2(\theta),
\end{align}
where we have introduced
\begin{align}
    k_{x'}(\theta) &= k_x\cos\theta - k_y \sin \theta, \notag \\
    k_{y'}(\theta) &= k_x \sin \theta + k_y \cos \theta.
\end{align}
\vspace{0.3cm}

\subsubsection{Scattering wave solutions}
Introducing a rectangular potential barrier of height $V_0$ between $x=0$ and $x=W$, we seek to find the scattering wave solutions for the Hamiltonian $\hat{\mathcal{H}}$ given by
\begin{align}\label{Equation General Elliptic Dirac Hamiltonian}
    \hat{\mathcal{H}} &= v_x \hat{p}_{x'}\tau_x + v_y \hat{p}_{y'} \tau_y + V_0(x) \tau_0, \hspace{0.1cm}
\end{align}
with $\tau_0$ the $2\times 2$ identity matrix and where $V_0(x)$ is defined as
\begin{align}
    V_0(x) &=
    \begin{cases}
        0, \hspace{0.1cm} \text{for} \hspace{0.1cm} x<0.\\
        V_0, \hspace{0.1cm} \text{for} \hspace{0.1cm} 0\leq x \leq W.\\
        0, \hspace{0.1cm} \text{for} \hspace{0.1cm} x>W.
    \end{cases}
\end{align}
To find these solutions, we first solve the time-independent Schr\"{o}dinger equation, $\hat{\mathcal{H}} \boldsymbol{\Psi} = E \boldsymbol{\Psi}$. Focusing first on the regions outside the potential barrier, we try a solution of the earlier suggested form,  $\boldsymbol{\Psi}_{\mathbf{k}}(x,y) = e^{i k_x x + i k_y y} \,\boldsymbol{\Psi}_0(\mathbf{k})$. We then arrive at the following algebraic equation relating the components of $\boldsymbol{\Psi}_{0}(\mathbf{k})$,
\begin{align}
    \hbar \bigg[ v_x k_{x'}(\theta) \tau_x  +  v_y k_{y'}(\theta) \tau_y  \bigg]\boldsymbol{\Psi}_0  =E \boldsymbol{\Psi}_0.
\end{align}
Recalling that the energy $E$ is dependent on the wave vector $\mathbf{k}$ via Eq.~\eqref{Equation Dirac Cone Energy}, and writing
\begin{align}
    \boldsymbol{\Psi}_0(\mathbf{k}) =
    \begin{pmatrix}
        \Psi_{+}(\mathbf{k}) \\
        \Psi_-(\mathbf{k})
    \end{pmatrix}
    ,
\end{align}
we find the following connection between the components of $\boldsymbol{\Psi}_{0}(\mathbf{k})$,
\begin{align}
    E\,\Psi_{-}(\mathbf{k}) &= |E|e^{i \phi(\mathbf{k})}\Psi_{+}(\mathbf{k}).
\end{align}

Here, the angle $\phi(\mathbf{k})$ is defined as
\begin{align}
    \phi(\mathbf{k}) = \arg \biggr[ v_x k_{x'}(\theta)+ i v_y k_{y'}(\theta)\biggr].
\end{align}
Given this connection, we then immediately find that the solutions to the time-independent Schr\"{o}dinger equation in the region outside the potential barrier are given by
\begin{align}
    \boldsymbol{\Psi}_{\mathbf{k}}(x,y) = e^{i k_x x + i k_y y}
    \begin{pmatrix}
        1\\
       s\, e^{i \phi(\mathbf{k})}
    \end{pmatrix}
    ,
\end{align}
with $s = \text{sgn}(E)$. In a similar fashion, the solution inside the potential barrier is found to be
\begin{align}
    \boldsymbol{\Psi}_{\mathbf{q}}(x,y) = e^{i q_x x + i q_y y}
    \begin{pmatrix}
        1\\
       s'\, e^{i \phi(\mathbf{q})}
    \end{pmatrix}
    ,
\end{align}
with $s' = \text{sgn}(E-V_0)$ and $\mathbf{q} = (q_x,q_y)$ any wave vector satisfying $E(\mathbf{q}) = E-V_0$, with $E(\mathbf{q})$ given in Eq.~\eqref{Equation Dirac Cone Energy}.

Having derived the energy eigenstates of $\hat{\mathcal{H}}$, we now construct the appropriate scattering wave solutions $\boldsymbol{\Psi}(x,y)$ of the system. Due to the translational invariance of the system along the $y$-direction, the $y$-component of the electron momentum is a conserved quantity. Hence, we find that the scattering wave solution for an incoming electron with momentum $\hbar \mathbf{k}$ and energy $E$ is given by
\begin{widetext}
\begin{align}\label{Equation Scattering Wave Solution}
    \boldsymbol{\Psi}(x,y) &=
    \begin{cases} 
    & \hspace{0.3cm}
        e^{i k_x x+i k_y y} 
        \begin{pmatrix}
            1\\
            s\,e^{ i \phi(\mathbf{k})}
        \end{pmatrix}
        +
        r\,e^{i k_r x+i k_y y} 
        \begin{pmatrix}
            1\\
            s\,e^{ i \phi(\mathbf{k}_r)}
        \end{pmatrix}
        ,\hspace{0.5cm} \text{for} \hspace{0.1cm} x< 0,\rule{0pt}{5ex}\\
        &\alpha \,e^{i q_x x+ i k_y y} 
        \begin{pmatrix}
            1\\
            s'\,e^{ i \phi(\mathbf{q})}
        \end{pmatrix}
        +
        \beta\,e^{i q_r x+i k_y y} 
        \begin{pmatrix}
            1\\
            s'\,e^{ i \phi(\mathbf{q}_r)}
        \end{pmatrix}
        ,\hspace{0.3cm} \text{for} \hspace{0.1cm} 0 \leq x \leq W,\rule{0pt}{5ex}\\
        &
        t \, e^{i k_x x+ik_y y} 
        \begin{pmatrix}
            1\\
            s\,e^{ i \phi(\mathbf{k})}
        \end{pmatrix}
        ,\hspace{4.6cm} \text{for} \hspace{0.1cm} x> 0.\rule{0pt}{5ex}
    \end{cases}
\end{align}
\end{widetext}

Here, $r,t \in \mathbb{C}$ are the reflection and transmission amplitudes, respectively, while $\alpha,\beta \in \mathbb{C}$ are complex coefficients defined in the barrier region. We have also introduced the reflected wave vector $\mathbf{k}_r = (k_r,k_y)$ and the reflected barrier wave vector $\mathbf{q}_r = (q_r,k_y)$. Moreover, since $k_y$ is a conserved quantity, we have $\mathbf{q}=(q_x, k_y)$. The components $k_r$ and $q_r$ of the reflected waves are found by solving the equations $E(k_r,k_y) = E$ and $E(q_r,k_y) = E-V_0$, respectively. Defining $\eta(k_y)$ as
\begin{align}
    \eta(k_y) &= \frac{\frac{1}{2}\sin(2\theta)(v_x^2-v_y^2) k_y }{v_x^2 \cos^2(\theta)+v_y^2 \sin^2(\theta)},
\end{align}
we have
\begin{align}
    k_r &= -k_x + 2 \eta(k_y),\\
    q_r &= -q_x+2\eta(k_y).
\end{align}
An expression for $q_x$ in terms of $E$ and $k_y$ is obtained in a similar fashion, and we find
\begin{align}
    q_x = \eta(k_y) + q(E,k_y),
\end{align}
where

\begin{align}
    q(E,k_y) &= \frac{\sqrt{\bigg(\frac{E-V_0}{\hbar}\bigg)^2\bigg[ v_x^2 \cos^2(\theta)+v_y^2 \sin^2(\theta) \bigg]-v_x^2 v_y^2 k_y^2}}{v_x^2 \cos^2(\theta)+v_y^2 \sin^2(\theta)}.
\end{align}

\begin{figure*}
    \centering
    \includegraphics[width=\linewidth]{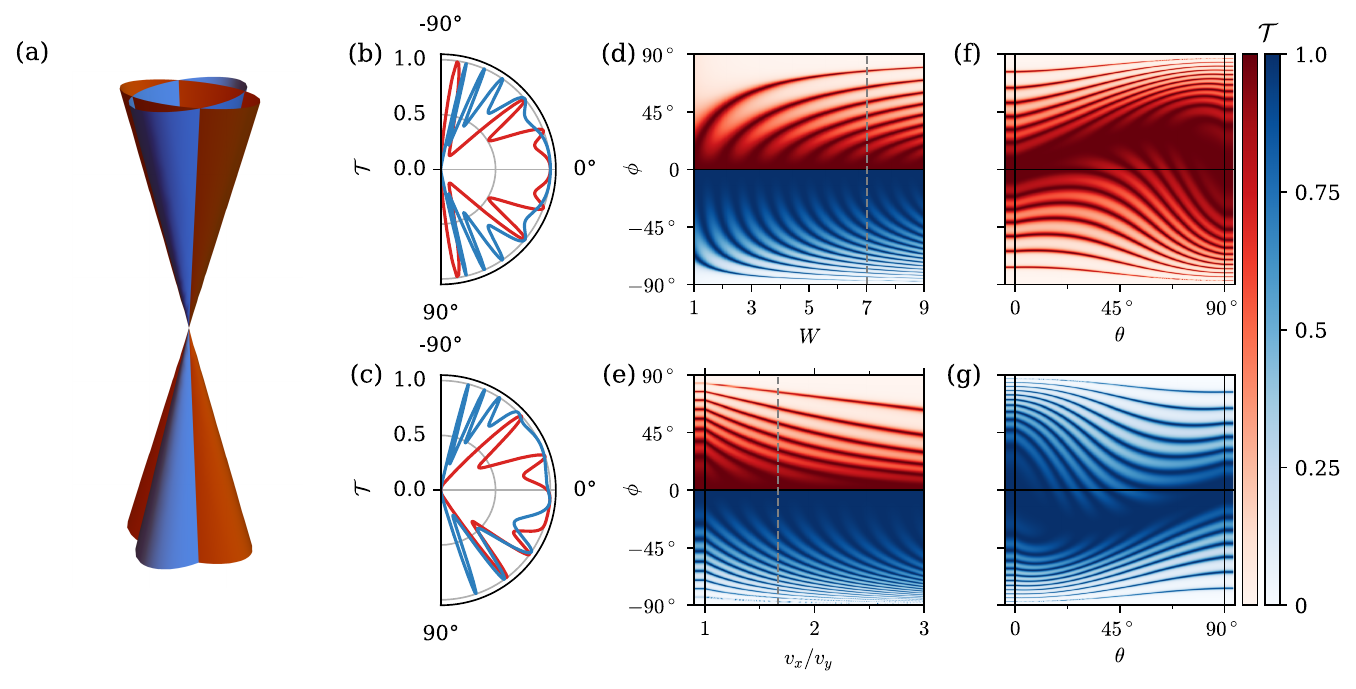}
    \caption{\textbf{Transmission of a $d$-wave Dirac altermagnet.}
    (a) Low-energy structure of a $d$-wave Dirac altermagnet.
    (b) Spin- and angle-resolved transmission, as a function of the group velocity angle $\varphi=\arctan(v^g_{y,\sigma}/v^g_{x,\sigma})$ of the incident electron, for an angle $\theta=0^\circ$ and (c) an angle $\theta = 18^\circ$ between the barrier and the altermagnet. Red (blue) denotes spin-up (spin-down) transmission. 
    (d) Spin- and angle-resolved transmission as a function of the barrier width $W$; the upper (lower) part of the panel corresponds to spin-up (spin-down).
    (e) Transmission as a function of the velocity anisotropy ratio $v_x/v_y$. 
    (f) Spin-up and (g) spin-down transmission as a function of the barrier-region orientation angle $\theta$. 
    Dashed lines in (d) and (e) indicate the parameters for which (b) and (c) are obtained. In all plots, results are shown for $V_0=10$ (arb. units), $E=0.4 \, V_0$, $W=7$ (arb. units), $v_F=1$ (arb. units), $\xi=0.25\, v_F$, and $\theta = 0^{\circ}$,  unless stated otherwise.}
    \label{fig:dwavetransmission}
\end{figure*}

\begin{figure*}
    \centering
    \includegraphics[width=\linewidth]{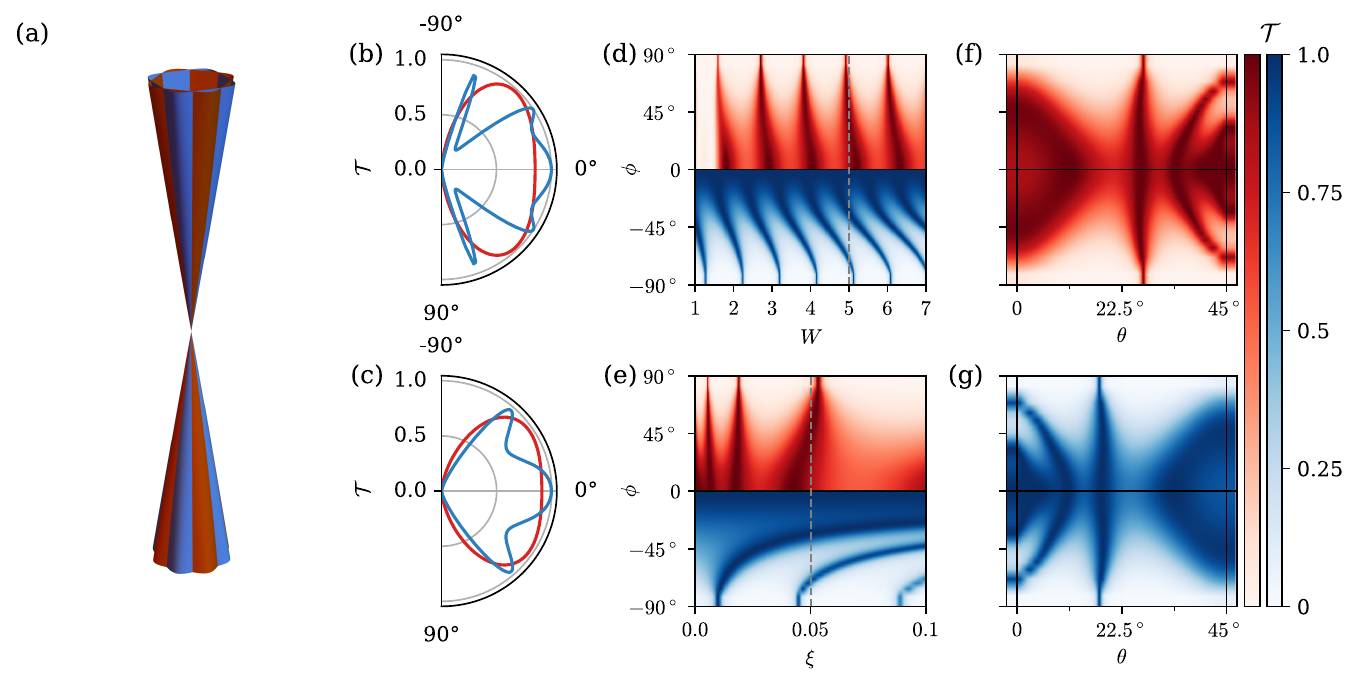}
    \caption{\textbf{Transmission of a $g$-wave Dirac altermagnet.} 
    (a) Low-energy structure of a $g$-wave Dirac altermagnet.
    Spin- and angle-resolved transmission, as a function of the group velocity angle $\varphi=\arctan(v^g_{y,\sigma}/v^g_{x,\sigma})$ of the incident electron, for an angle $\theta=0^\circ$ and (c) an angle $\theta = 5.8^\circ$ between the barrier and the altermagnet. Red (blue) denotes spin-up (spin-down) transmission. 
    (d) Spin- and angle-resolved transmission as a function of the barrier width $W$; the upper (lower) part of the panel corresponds to spin-up (spin-down).
    (e) Transmission as a function of the form factor constant $\xi$. 
    (f) Spin-up and (g) spin-down transmission as a function of the barrier-region orientation angle $\theta$. 
    Dashed lines in (d) and (e) indicate the parameters for which (b) and (c) are obtained. In all plots, results are shown for
    $V_0=5$ (arb. units), $E=0.2 \, V_0$, $W=5$ (arb. units), $v_F=1$ (arb. units), $\xi=0.05\, v_F$, and $\theta = 0^{\circ}$,  unless stated otherwise.}
    \label{fig:gwavetransmission}
\end{figure*}

\subsubsection{Transmission probability}
Having derived the scattering wave solutions $\boldsymbol{\Psi}(x,y)$ in Eq.~\eqref{Equation Scattering Wave Solution}, we now seek to determine the transmission amplitude $t$ and the transmission probability $\mathcal{T} = |t|^2$. This can be done by demanding continuity of the wave function at the boundaries of the potential barrier,
\begin{align}
    \lim_{x \rightarrow 0^-} \boldsymbol{\Psi}(x,y) &= \lim_{x \rightarrow 0^+} \boldsymbol{\Psi}(x,y),\\
    \lim_{x \rightarrow W^-} \boldsymbol{\Psi}(x,y) &= \lim_{x \rightarrow W^+} \boldsymbol{\Psi}(x,y),
\end{align}
which yields a system of linear equations from which the transmission amplitude can be calculated. The details of this calculation, as well as the most general result for $t$ that is valid for arbitrary $\theta$, are given in Appendix \ref{Appendix Transmission Amplitude}. Here, to obtain an intuitive understanding regarding the fundamental properties of the transmission amplitude, we simply state the result valid for the Klein tunneling limit, $|V_0| \gg |E|$, with $\theta =0$. We have
\begin{align}
    t(\mathbf{k},E) \approx \frac{ e^{-i k_x W} \cos \phi_0(\mathbf{k})}{\cos(q_0W)\cos \phi_0(\mathbf{k})-i s s' \sin(q_0W)},
\end{align}
where $\phi_0(\mathbf{k})$ and $q_0$ are given by
\begin{align}
    \phi_0(\mathbf{k}) &= \arg(v_x k_x + i v_y k_y),\\
    q_0 &= \frac{1}{v_x} \sqrt{\biggr(\frac{E-V_0}{\hbar}\biggr)^2-v_y^2k_y^2}.
\end{align}
The corresponding transmission coefficient $\mathcal{T}$ is then immediately found to be equal to
\begin{align}\label{Equation Tranmission Coefficient Unrotated Elliptic Dirac Cone}
    \mathcal{T}(\mathbf{k},E) \approx \frac{\cos^2 \phi_0(\mathbf{k})}{1- \cos^2(q_0W)\sin^2\phi_0(\mathbf{k})}.
\end{align}
As expected, taking $v_x = v_y$, we obtain the graphene transmission coefficient derived earlier by Katsnelson \textit{et al.} \cite{Katsnelson2006}. Several of its important properties are retained in the case where $v_x \neq v_y$, the most prominent example being perfect transmission ($\mathcal{T} = 1$) at normal incidence, the defining characteristic of Klein tunneling. An important difference, however, is that the angle governing the transmission is no longer the angle between $\mathbf{k}$ and the normal to the potential barrier, like in graphene, but rather the angle between $v_x k_x \hat{\mathbf{x}}+v_y k_y \hat{\mathbf{y}}$ and said normal vector. Finally, while rotations by an angle $\theta$ do not affect Klein tunneling in graphene, they can significantly impact the transmission for elliptic Dirac cones.

\subsection{Results}

Plots of the spin-dependent transmission coefficient $\mathcal{T}_{\sigma}$ for the $d$-wave Dirac altermagnet are given in Fig.~\ref{fig:dwavetransmission}. Its low-energy structure is shown in Fig.~\ref{fig:dwavetransmission}(a), and it consists of two equivalent elliptic Dirac cones - one for each spin species - rotated by $90^{\circ}$ relative to one another. In Figs.~\ref{fig:dwavetransmission}(b) and (c), we plot $\mathcal{T}_{\sigma}$ for $\theta = 0^{\circ}$ and $\theta = 18^{\circ}$, respectively, as a function of the angle $\varphi(\mathbf{k}) = \arctan \left[v^g_{y,\sigma}(\mathbf{k})/v^g_{x,\sigma}(\mathbf{k})\right]$, where $v^g_{j,\sigma}$ ($j=x,y$) is the $j$-component of the group velocity $\mathbf{v}^g_{\sigma}(\mathbf{k}) = (1/\hbar)\nabla_{\mathbf{k}}E_{\sigma}(\mathbf{k})$. We observe a pronounced spin dependence of the transmission, both in the angular profile and in the position of the transmission maxima. Most notably, for $\theta = 0^{\circ}$, we find that perfect transmission occurs at normal incidence ($\varphi = 0$) for both spin species, while for $\theta = 18^{\circ}$, the broadest perfect-transmission peak shifts away from $\varphi = 0$, a clear signature of anomalous Klein tunneling \cite{Bentancur-Ocampo2018,Iurov2020,Betancur-Ocampo2026}, and occurs at different angles for the two spin species. The other peaks in the transmission correspond to Fabry-P\'{e}rot resonances, and they signify the strong dependence of $\mathcal{T}_{\sigma}$ on the barrier width $W$ and the velocity anisotropy ratio $v_x/v_y$, as illustrated in Figs.~\ref{fig:dwavetransmission}(d) and (e). While the latter is a material parameter that is typically challenging to alter, the barrier width can be modified rather easily. In fact, a closer inspection of Eq.~\eqref{Equation Tranmission Coefficient Unrotated Elliptic Dirac Cone} reveals that the transmission coefficient depends directly on $q_0 W$, and therefore we can also alter $\mathcal{T}_{\sigma}$ by changing the barrier height $V_0$ while keeping $W$ fixed. As a minor comment, we mention that for $\theta =0^\circ$, $\mathcal{T}_{\sigma}$ is symmetric under the transformation $\varphi \rightarrow - \varphi$, which allows us to fully characterize $\mathcal{T}_{\uparrow}$ ($\mathcal{T}_{\downarrow}$) by focusing on positive (negative) angles $\varphi$ ranging from $0^{\circ}$ to $90^{\circ}$ ($-90^{\circ}$). Finally, in  Figs.~\ref{fig:dwavetransmission}(f) and (g), we demonstrate the strong dependence of the transmission coefficient on the barrier orientation angle $\theta$, thus introducing yet another tuning knob to control Klein tunneling in this system. We note that $\mathcal{T}_{\uparrow}$ and $\mathcal{T}_{\downarrow}$ in these two plots are related to each other via a $90^{\circ}$ rotation, as one would intuitively expect from the rotational symmetries of the $d$-wave altermagnet, and that the anomalous Klein tunneling effect is again visibly present.

For the $g$-wave altermagnet, we consider the Hamiltonian
\begin{align}
    \hat{\mathcal{H}} &= v_F \, \sigma_0(\mathbf{p} \cdot \boldsymbol{\tau}) + \xi  \begin{pmatrix}
        (p_x^2-p_y^2)^2 & 0 \\
        0 & 4p_x^2 p_y^2 
    \end{pmatrix} \tau_z\label{eq:GwaveH} .
\end{align}
While an analytical calculation of the transmission coefficient for the $g$-wave Dirac altermagnet is not feasible, a numerical approach along a similar line of reasoning is readily available. The main complications here are the fact that the time-independent Schr\"{o}dinger equation now becomes a fourth-order equation in $k_x$ and that we have to take into account evanescent modes as well. For a detailed discussion, we refer the reader to Appendix \ref{Appendix G-Wave}.

Plots of the spin-dependent transmission coefficient $\mathcal{T}_{\sigma}$ for the $g$-wave Dirac altermagnet are given in Figure \ref{fig:gwavetransmission}. Its low-energy structure is presented in Fig.~\ref{fig:gwavetransmission}(a), while in Figs.~\ref{fig:gwavetransmission}(b) and (c), we plot $\mathcal{T}_{\sigma}$ for $\theta = 0^{\circ}$ and $\theta = 5.8^{\circ}$, respectively, as a function of the angle $\varphi(\mathbf{k}) = \arctan \left[v^g_{y,\sigma}(\mathbf{k})/v^g_{x,\sigma}(\mathbf{k})\right]$. Like before, a pronounced spin dependence is observed in the transmission. However, due to the presence of higher-order momentum terms in the Hamiltonian, we do not always find perfect transmission at normal incidence. In fact, in both figures we see that the spin-up electrons are never perfectly transmitted for the chosen parameters, with the near-unity transmission at $\theta = 0^{\circ}$ arising only because we are close to Fabry-P\'erot resonance values for the width $W$ and the form factor constant $\xi$, as can be seen from Figs.~\ref{fig:gwavetransmission}(d) and (e). In contrast, the spin-down electrons do exhibit Klein tunneling at both angles. As further shown in Figs.~\ref{fig:gwavetransmission}(f) and (g), the emergence of the Klein tunneling regime depends sensitively on the barrier orientation angle $\theta$, with Klein tunneling occuring for spin-down electrons when $\theta$ is close to zero, whereas for spin-up electrons it appears when $\theta$ approaches to $45^{\circ}$. Finally, we notice that the rotational symmetries of the system allow us to fully characterize $\mathcal{T}_{\uparrow}$ ($\mathcal{T}_{\downarrow}$) by focusing on positive (negative) angles $\varphi$ ranging from $0^{\circ}$ to $45^{\circ}$ ($-45^{\circ}$).

\section{Spin Currents}\label{Section Spin Currents}
Having determined the transmission coefficients for $\ell$-wave Dirac altermagnets, we now proceed to calculate the electronic spin currents using the Landauer-B\"{u}ttiker formalism \cite{Landauer1957,Buttiker1985,Buttiker1986}. We first derive the resulting integral expressions valid for any $\ell$-wave Dirac altermagnet, before specializing to the case $\ell = d$, which can again be tackled by considering a single elliptic Dirac band. Finally, we present plots of the spin-current polarization as a function of parameters such as the barrier height and barrier orientation.

\subsection{Spin currents for $\ell$-wave Dirac altermagnets}
Denoting the regions to the left and right of the potential barrier by $L$ and $R$, respectively, we take them to be incoherent electron reservoirs with temperatures $T_L$ and $T_R$, and chemical potentials $\mu_L$ and $\mu_R$, respectively. We note that the incoherent reservoir assumption in the Landauer-B\"{u}ttiker formalism allows one to ignore interference effects between electrons emitted from the reservoir $L$ with electrons emitted from the reservoir $R$. Electron transport through the potential barrier is assumed to be phase-coherent, and the barrier width $W$ is therefore required to be much smaller than the phase coherence length. On the other hand, in the direction transverse to the barrier normal, we take the system size to be much larger than the phase coherence length, so that transverse mode quantization can be neglected.

Assuming that a small voltage difference $V$ is applied between the reservoirs, a current $I$ will flow through the barrier. In principle, this bias modifies the potential profile inside the barrier and could be modelled by a spatially varying (e.g., linear) potential. However, for sufficiently small $V$, these modifications can be neglected and the transport can be described using spin-dependent zero-bias transmission coefficients $\mathcal{T}_{\sigma}(\mathbf{k},E)$ .

Following the Landauer-B\"{u}ttiker formalism, we then find that the current density flowing from reservoir $L$ to reservoir $R$ is given by
\begin{align}
    \mathbf{J}_{L\rightarrow R,\sigma} = e\sum_{\gamma=\pm} \int_{\Omega_{+\sigma}} &\frac{d^2 \mathbf{k}}{(2\pi)^2} \, \mathbf{v}^g_{\sigma}(\mathbf{k},E_{\gamma\sigma}) \mathcal{T}_{\sigma}(\mathbf{k},E_{\gamma\sigma}) \nonumber\\
    &\times f_{L}(E_{\gamma\sigma})[1-f_{R}(E_{\gamma\sigma})],
\end{align}
while the current density flowing from reservoir $R$ to reservoir $L$ is given by
\begin{align}
    \mathbf{J}_{R\rightarrow L,\sigma} = e\sum_{\gamma=\pm} \int_{\Omega_{-\sigma}} &\frac{d^2 \mathbf{k}}{(2\pi)^2} \, \mathbf{v}^g_{\sigma}(\mathbf{k},E_{\gamma\sigma}) \mathcal{T}_{\sigma}(\mathbf{k},E_{\gamma\sigma}) \nonumber\\
    &\times f_{R}(E_{\gamma\sigma})[1-f_{L}(E_{\gamma\sigma})].
\end{align}
Here, $\mathbf{v}^g_{\sigma} = (1/\hbar) \nabla_{\mathbf{k}}E_{\sigma}$ denotes the spin-dependent group velocity, while $E_{\gamma\sigma}(\mathbf{k})$ (with $\gamma=\pm$) corresponds to the positive and negative spin-dependent energies of the Dirac altermagnets, respectively. Furthermore, $e$ is the electron charge, and we have introduced the Fermi-Dirac distributions $f_{L/R}$ of the reservoirs $L$ and $R$, where 
\begin{align}
    f_{L}(E) &= \frac{1}{1+e^{(E-\mu_{L})/k_B T_{L}}},\\
    f_{R}(E) &= \frac{1}{1+e^{(E-\mu_{R})/k_B T_{R}}}.
\end{align}
In what follows, we take the reservoir temperatures to be the same, $T_L = T_R = T$, and the chemical potentials to be given by $\mu_{L} = eV$ and $\mu_R = 0$. Finally, the total spin current is given by $\mathbf{J}_{\sigma} = \mathbf{J}_{L \rightarrow R,\sigma}+\mathbf{J}_{R \rightarrow L,\sigma}$, and the spin-dependent integration domain $\Omega_{+\sigma}$ ($\Omega_{-\sigma}$) corresponds to all $\mathbf{k}$ for which $v^g_{x,\sigma}(\mathbf{k},E_{\sigma})$ is a positive (negative) number.

Now, to be useful for spintronics applications, we require that there is an imbalance between the spin currents, meaning $\mathbf{J}_{\uparrow} \neq \mathbf{J}_{\downarrow}$. As the voltage difference is applied along $x$, an appropriate quantitative measure for this imbalance is the spin-current polarization $\mathcal{P}$ of the $x$-component of the current, which in our case is given by
\begin{align}\label{Equation Spin-Current polarization Definition}
    \mathcal{P} = \frac{J_{x,\uparrow}-J_{x,\downarrow}}{J_{x,\uparrow}+J_{x,\downarrow}}.
\end{align}
The ability to control $\mathcal{P}$ is therefore of significant importance, and we will demonstrate in Section \ref{Section Spin Currents Results} that Klein tunneling can be used to both significantly enhance and diminish the spin-current polarization.

\subsection{Current for a single elliptic Dirac Hamiltonian}
To understand the properties of the spin currents in Dirac $d$-wave altermagnets, we can again focus our attention on a system described by the (sub-)Hamiltonian corresponding to $\sigma = \uparrow$,
\begin{align}
    \hat{\mathcal{H}}_{\uparrow} &= v_x \hat{p}_{x'}\tau_x + v_y \hat{p}_{y'} \tau_y + V_0(x) \tau_0. 
\end{align}
The resulting transmission coefficient $\mathcal{T}_{\uparrow}(\mathbf{k})$ is calculated in Appendix \ref{Appendix Transmission Amplitude}, and the group velocity $\mathbf{v}^g_{\uparrow}(\mathbf{k},E_{\uparrow})$, 
\begin{align}
    \mathbf{v}^{g}_{\uparrow}(\mathbf{k},E_{\uparrow}) &= v^g_{x,\uparrow}(\mathbf{k},E_{\uparrow}) \hat{\mathbf{x}}+v^g_{y,\uparrow}(\mathbf{k},E_{\uparrow}) \hat{\mathbf{y}},
\end{align}
has components $v^g_{x,\uparrow}$ and $v^g_{y,\uparrow}$ that are given by
\begin{align}
v^g_{x,\uparrow}(\mathbf{k},E_{\uparrow})
&= \frac{\hbar}{E_{\uparrow}}\Big[
 (v_x^2 \cos^2\theta + v_y^2 \sin^2\theta)k_x
\nonumber\\
&\qquad
 + \tfrac12 \sin(2\theta)(v_y^2 - v_x^2)k_y
\Big], \\
v^g_{y,\uparrow}(\mathbf{k},E_{\uparrow})
&= \frac{\hbar}{E_{\uparrow}}\Big[
 (v_x^2 \sin^2\theta + v_y^2 \cos^2\theta)k_y
\nonumber\\
&\qquad
 + \tfrac12 \sin(2\theta)(v_y^2 - v_x^2)k_x
\Big].
\end{align}

\begin{figure*}
    \centering
    \includegraphics[width=\textwidth]{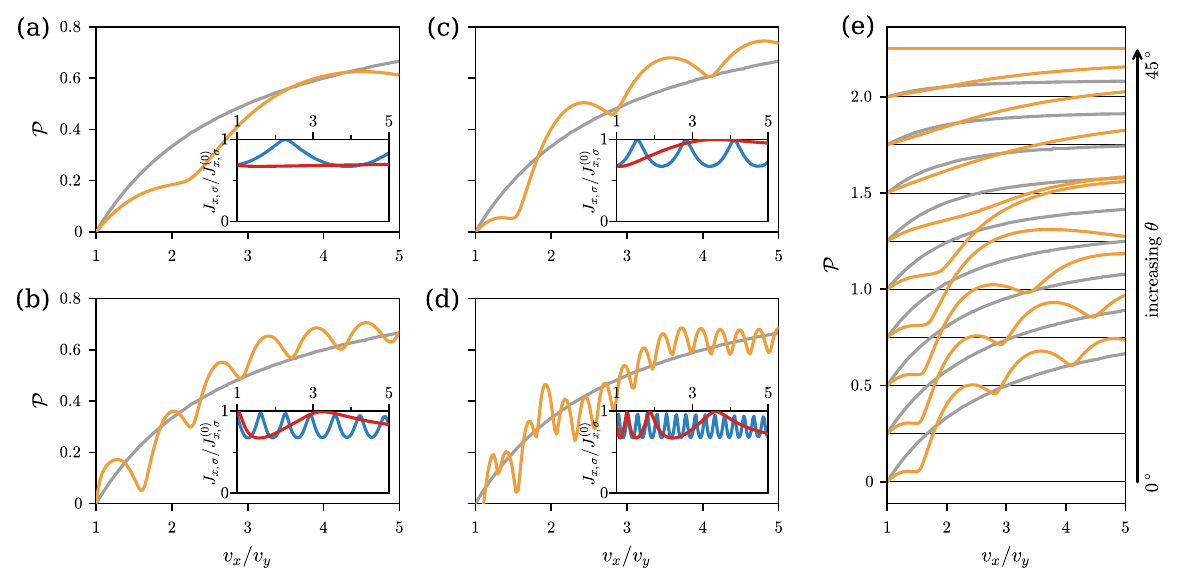}
    \caption{
    \textbf{Spin-current polarization $\mathcal{P}$ in a $d$-wave Dirac altermagnet.}
    Yellow curves show the spin-current polarization $\mathcal{P}$ for different barrier parameters: 
    (a) $V_0 = 20$ (arb. units), $eV = 0.05 \,V_0$, $W = 0.1$ (arb. units); 
    (b) $V_0 = 20$ (arb. units), $eV = 0.05 \,V_0$, $W = 0.5$ (arb. units); 
    (c) $V_0 = 50$ (arb. units), $eV = 0.02 \,V_0$, $W = 0.1$ (arb. units); 
    (d) $V_0 = 50$ (arb. units), $eV = 0.02 \,V_0$, $W = 0.5$ (arb. units),
    as a function of the velocity anisotropy ratio $v_x/v_y$.
    The gray line indicates the spin-current polarization in the absence of a barrier, i.e., the intrinsic polarization of the Dirac altermagnet. 
    Insets show the spin-resolved current ratio between the barrier ($J_{x,\sigma}$) and no-barrier ($J^{(0)}_{x,\sigma}$) cases, with red for spin-up electrons and blue for spin-down electrons, and we have used barrier angle $\theta = 0^{\circ}$. 
    (e) Spin-current polarization corresponding to the parameters of panel (c), shown as a function of the barrier angle $\theta$ ranging from $0^\circ$ to $45^\circ$. For clarity, the curves for different $\theta$ values are vertically offset. The spacing between the black horizontal lines corresponds to $\Delta \mathcal{P} = 0.25$, and each successive curve represents an increase of $5^\circ$ in $\theta$.
     In all plots, results are shown for $v_F=1$ (arb. units) and $T = 0$.
    }
    \label{fig:curdwave}
\end{figure*}

\begin{figure*}
    \centering
    \includegraphics[width=\textwidth]{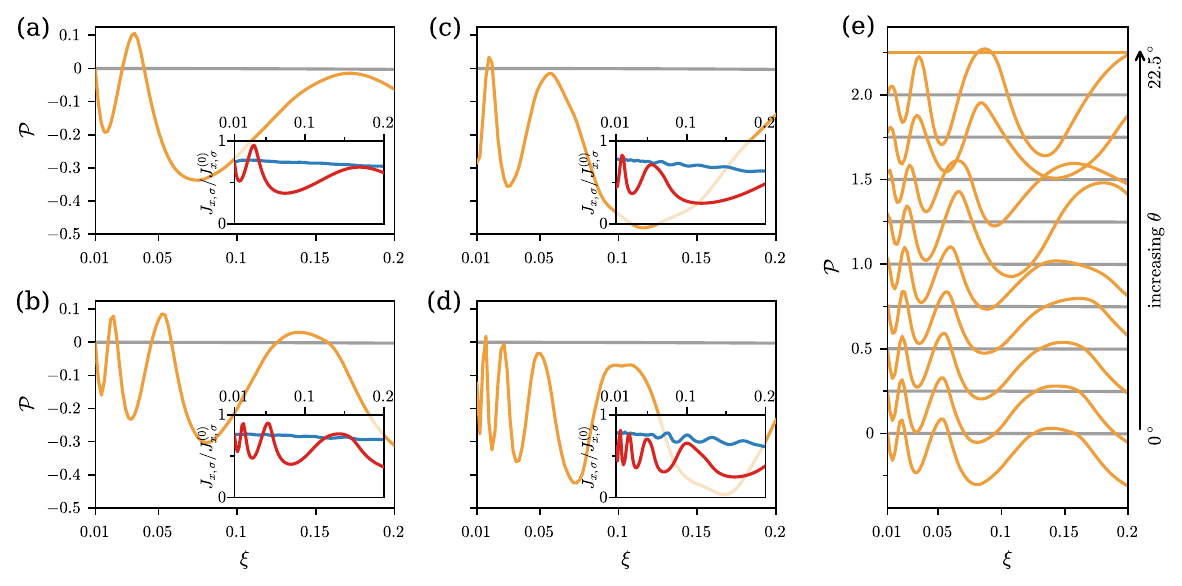}
    \caption{
    \textbf{Spin-current polarization $\mathcal{P}$ in a $g$-wave Dirac altermagnet.}
    Yellow curves show the spin-current polarization $\mathcal{P}$ for different barrier parameters: 
    (a) $V_0 = 5$ (arb. units), $eV = 0.2 \,V_0$, $W = 2.5$ (arb. units); 
    (b) $V_0 = 5$ (arb. units), $eV = 0.2 \,V_0$, $W = 5$ (arb. units); 
    (c) $V_0 = 10$ (arb. units), $eV = 0.1 \,V_0$, $W = 2.5$ (arb. units); 
    (d) $V_0 = 10$ (arb. units), $eV = 0.1 \,V_0$, $W = 5$ (arb. units),
    as a function of the form factor constant $\xi$.
    The gray line indicates the spin-current polarization in the absence of a barrier, i.e., the intrinsic polarization of the Dirac altermagnet. 
    Insets show the spin-resolved current ratio between the barrier ($J_{x,\sigma}$) and no-barrier ($J^{(0)}_{x,\sigma}$) cases, with red for spin-up electrons and blue for spin-down electrons, and we have used barrier angle $\theta = 0^{\circ}$. 
    (e) Spin-current polarization corresponding to the parameters of panel (c), shown as a function of the barrier angle $\theta$ ranging from $0^\circ$ to $45^\circ$. For clarity, the curves for different $\theta$ values are vertically offset. The spacing between the black horizontal lines corresponds to $\Delta \mathcal{P} = 0.25$, and each successive curve represents an increase of $2.5^\circ$ in $\theta$.
     In all plots, results are shown for $v_F=1$ (arb. units) and $T = 0$.
    }
    \label{fig:curgwave}
\end{figure*}

Although we are now in principle able to calculate the total spin-up current density  $\mathbf{J}_{\uparrow} = \mathbf{J}_{R \rightarrow L,\uparrow}+\mathbf{J}_{L \rightarrow R,\uparrow}$, it is often convenient to change integration coordinates from $(k_x,k_y)$ to $(E_{\uparrow},\zeta)$. Here, $\zeta$ is the angle between $\mathbf{k}$ and the vector $\hat{\mathbf{x}}' = \cos \theta \, \hat{\mathbf{x}}+\sin \theta \, \hat{\mathbf{y}}$. Introducing the Heaviside step function $H$, we then immediately have
\begin{align}
\mathbf{J}_{L\rightarrow R,\uparrow}
&= \frac{e}{(2\pi)^2}
\sum_{\gamma=\pm} \gamma
\int_0^{\gamma \infty}\!\! dE_{\uparrow}
\int_0^{2\pi}\!\! d\zeta \bigg| \frac{\partial (k_x,k_y)}{\partial (E_{\uparrow},\zeta)} \bigg|\nonumber\\
&\quad \times 
\, \mathbf{v}^g_{\uparrow}(\zeta,E_{\uparrow})
\, \mathcal{T}_{\uparrow}(\zeta,E_{\uparrow})
\, f_L(E_{\uparrow})[1-f_R(E_{\uparrow})] \nonumber\\
&\quad \times
H\!\left(v^g_{x,\uparrow}(E_{\uparrow},\zeta)\right),
\\[6pt]
\mathbf{J}_{R\rightarrow L,\uparrow}
&= \frac{e}{(2\pi)^2}
\sum_{\gamma=\pm} \gamma
\int_0^{\gamma \infty}\!\! dE_{\uparrow}
\int_0^{2\pi}\!\! d\zeta \bigg| \frac{\partial (k_x,k_y)}{\partial (E_{\uparrow},\zeta)} \bigg| \nonumber\\
&\quad \times 
\, \mathbf{v}^g_{\uparrow}(\zeta,E_{\uparrow})
\, \mathcal{T}_{\uparrow}(\zeta,E_{\uparrow})
\, f_R(E_{\uparrow})[1-f_L(E_{\uparrow})] \nonumber\\
&\quad \times
H\!\left(-v^g_{x,\uparrow}(E_{\uparrow},\zeta)\right).
\end{align}

Here, we have introduced the Jacobian,
\begin{align}
    \biggr( \frac{\partial k_x, \partial k_y}{\partial E_{\uparrow}, \partial \zeta}\biggr) &= 
    \det
    \begin{pmatrix}
        \frac{\partial k_x}{\partial E_{\uparrow}} & \frac{\partial k_x}{\partial \zeta}\\
        \frac{\partial k_y}{\partial E_{\uparrow}} & \frac{\partial k_y}{\partial \zeta}
    \end{pmatrix}
    ,
\end{align}
and we have also defined $\mathbf{v}^g_{\uparrow}(\zeta,E_{\uparrow})=\mathbf{v}^g_{\uparrow}(\mathbf{k},E_{\uparrow})$ and $\mathcal{T}_{\uparrow}(\zeta,E_{\uparrow})=\mathcal{T}_{\uparrow}(\mathbf{k},E_{\uparrow})$, with $\mathbf{k} = \mathbf{k}(E_{\uparrow},\zeta)$. In the simple case where $\theta = 0$, one can immediately write down 
\begin{align}
    \mathbf{k}(E_{\uparrow},\zeta) = \biggr|\frac{E_{\uparrow}}{\hbar v_x}\biggr| \cos \zeta \, \hat{\mathbf{x}}+\biggr|\frac{E_{\uparrow}}{\hbar v_y}\biggr| \sin \zeta \, \hat{\mathbf{y}},
\end{align}
and we then get
\begin{align}
    \mathbf{J}_{L\rightarrow R,\uparrow} &= \frac{e}{h^2 |v_y| } \int_{-\infty}^{\infty} d E \int_{-\frac{\pi}{2}}^{\frac{\pi}{2}} d \zeta\\
    &\times \frac{|E|\,\cos^3(\zeta)\,\hat{\mathbf{x}}}{1-\cos^2(q_{0,\uparrow}(E)W) \sin^2(\zeta)} f_L(E)[1-f_R(E)], \nonumber\\
    \mathbf{J}_{R\rightarrow L,\uparrow} &= \frac{-e}{h^2 |v_y| } \int_{-\infty}^{\infty} d E \int_{-\frac{\pi}{2}}^{\frac{\pi}{2}} d \zeta \\
    &\times \frac{|E|\,\cos^3(\zeta)\,\hat{\mathbf{x}}}{1-\cos^2(q_{0,\uparrow}(E)W) \sin^2(\zeta)} f_R(E)[1-f_L(E)]. \nonumber
\end{align}
We note that there are no further analytical simplifications possible, and one typically has to resort to numerical integration techniques to evaluate $\mathbf{J}_{\uparrow}$.

Finally, we observe that, regardless of the values of $\theta$, we can directly use the above results to obtain the spin-down current $\mathbf{J}_{\downarrow}$ for a $d$-wave Dirac altermagnet, since we only have to make the replacements $v_{x} \rightarrow v_y$ and $v_y \rightarrow v_x$.

\subsection{Results}\label{Section Spin Currents Results}

Plots of the spin-current polarization, as defined in Eq.~\eqref{Equation Spin-Current polarization Definition}, for the $d$-wave Dirac altermagnet are given in Fig.~\ref{fig:curdwave}. In Figs.~\ref{fig:curdwave}(a)–(d), we show $\mathcal{P}$ as a function of the velocity anisotropy ratio $v_x/v_y$ for different barrier heights and widths. The gray curve denotes the intrinsic polarization in the absence of a barrier, while the yellow curves show the polarization in the presence of a finite barrier. We find that the barrier can substantially modify the spin-current polarization. Depending on the barrier parameters, $\mathcal{P}$ may either be enhanced or suppressed relative to the intrinsic value due to the altermagnet, demonstrating that the barrier acts as an effective tuning knob for the spin polarization.

This behavior originates from the spin-dependent transmission probabilities associated with Klein tunneling in the anisotropic Dirac spectrum. As illustrated in the insets of Figs.~\ref{fig:curdwave}(a)–(d), the spin-resolved currents $J_{x,\sigma}$ remain sizable even in the presence of the barrier, indicating that the enhancement of $\mathcal{P}$ is not accompanied by a strong suppression of the overall current. Instead, the barrier selectively modifies the transmission of the two spin channels, leading to a redistribution of the spin-resolved currents.

In Fig.~\ref{fig:curdwave}(e) we further explore the role of the barrier orientation by plotting $\mathcal{P}$ as a function of $v_x/v_y$ for different barrier angles $\theta$. The curves are vertically offset for clarity. The results show that the barrier angle provides an additional degree of control over the spin polarization, allowing for continuous tuning of $\mathcal{P}$. Finally, we observe that $\mathcal{P}$ vanishes at $\theta = 45^{\circ}$ both with and without a barrier. This follows from the spinful mirror-$x$ symmetry relation $E_{\uparrow}(\mathbf{k}) = E_{\downarrow}(\mathbf{k})$, which enforces identical dispersions for the two spin channels at this orientation and therefore eliminates the net spin polarization.

Plots of the spin-current polarization for the $g$-wave Dirac altermagnet are given in Fig.~\ref{fig:curgwave}. In Figs.~\ref{fig:curgwave}(a)-(d), we present plots of $\mathcal{P}$ as a function of the form factor constant $\xi$ for different values of the barrier height and width. Compared to the $d$-wave case, the barrier parameters used here are smaller, since a sufficiently large potential barrier would push the states inside the barrier far away from the Dirac point, thereby suppressing the Klein tunneling regime and leading instead to predominantly evanescent transmission. The values of $V_0$ considered here therefore ensure that propagating Dirac states remain available inside the barrier, allowing Klein tunneling to occur.

Although we have already seen that Klein tunneling can lead to a significant enhancement of the spin-current polarization in the $d$-wave case, this effect is even more pronounced for the $g$-wave Dirac altermagnet. For the small values of $\xi$ considered here, the intrinsic spin-current polarization is nearly zero, as indicated by the gray curves. However, the inclusion of a potential barrier can significantly enhance the modulus of $\mathcal{P}$, with the polarization increasing by several orders of magnitude in certain parameter regimes. As shown in the insets of Figs.~\ref{fig:curgwave}(a)-(d), sizable spin currents are still maintained, indicating that this enhancement is not accompanied by a strong suppression of the total transmitted current.

This behavior can be understood by revisiting the transmission coefficient plots in Fig.~\ref{fig:gwavetransmission}, which show that for certain ranges of the barrier angle $\theta$, only one spin species exhibits perfect transmission at normal incidence while the other spin channel is strongly suppressed. The barrier therefore acts as an efficient spin filter, leading to a large enhancement of the spin-current polarization.

In Fig.~\ref{fig:curgwave}(e), we present plots of $\mathcal{P}$ as a function of $\xi$ for different values of the barrier angle. The curves are vertically offset for clarity. As before, we observe that $\mathcal{P}$ goes to zero as $\theta$ approaches $22.5^{\circ}$, which reflects the symmetry point where the dispersions of the two spin channels become identical and the net spin-current polarization therefore vanishes.

Although we have here presented plots that were evaluated at temperature $T=0$, we have verified that the observed effects are retained in the case of finite temperatures. Despite the fact that we have mostly worked with arbitrary units, we can also straightforwardly provide some estimated values for the relevant parameters. Using Klein tunneling in graphene \cite{Stander2009,Young2009} as our reference point, we expect typical experimental values to be $V_0 = 20-200$ meV, $W = 50-500$ nm and $T = 5-80$ K. These parameter ranges are well within reach of current experimental techniques and have been routinely realized in modern graphene devices.

\subsection{Resonance conditions}
In the previous section, we found that the spin-current polarization $\mathcal{P}$ in $\ell$-wave Dirac altermagnets depends strongly on both the barrier height $V_0$ and the barrier width $W$. Ideally, one would like to be able to tune these parameters in such a way that one can either maximize or minimize the absolute value of $\mathcal{P}$. Although it is a challenging task to find the optimal parameters for the $g$-wave Dirac altermagnet, an analytical calculation can readily be performed for the $d$-wave case with $\theta=0$, as we will demonstrate here.

Assuming a potential difference $V>0$ \footnote{A similar argument can be constructed for a potential difference $V<0$.}, it is easily seen that $\mathcal{P}$ is a monotonically increasing (decreasing) function of the positive ratio $J_{x,\uparrow}/J_{x,\downarrow}$ ($J_{x,\downarrow}/J_{x,\uparrow}$). Since the dependence of the current on the barrier width $W$ is only contained within the transmission coefficients $\mathcal{T}_{\sigma}$, a necessary requirement for $\mathcal{P}$ to be maximal is then
\begin{align}\label{Equation Resonance Condition 1}
    \frac{\partial}{\partial W}\biggr( \frac{\mathcal{T}_{\uparrow}(\mathbf{k},E_{\uparrow})}{\mathcal{T}_{\downarrow}(\mathbf{k}',E_{\downarrow})}\biggr) &= 0 \hspace{0.1cm} \Rightarrow \frac{\sin(2 q_{0,\uparrow}W)}{\sin(2 q_{0,\downarrow}W)} = 0,
\end{align}
for all $\mathbf{k},\mathbf{k}' \in \mathbb{R}^2$, while the corresponding requirement for $\mathcal{P}$ to be minimal is given by
\begin{align}\label{Equation Resonance Condition 2}
    \frac{\partial}{\partial W}\biggr( \frac{\mathcal{T}_{\downarrow}(\mathbf{k},E_{\downarrow})}{\mathcal{T}_{\uparrow}(\mathbf{k}',E_{\uparrow})}\biggr) &= 0 \hspace{0.1cm} \Rightarrow \frac{\sin(2 q_{0,\downarrow}W)}{\sin(2 q_{0,\uparrow}W)} = 0.
\end{align}
To obtain the equations on the right-hand side of the arrow, we have made use of Eq.~\eqref{Equation Tranmission Coefficient Unrotated Elliptic Dirac Cone}, and we have introduced
\begin{align}
    q_{0,\uparrow} &= \frac{V_0}{\hbar v_x} \qquad \text{and} \qquad q_{0,\downarrow} = \frac{V_0}{\hbar v_y}.
\end{align}
Now, taking $n \in \mathbb{Z}$, we find that all solutions to Eq.~\eqref{Equation Resonance Condition 1} are of the form
\begin{align}
    W_{n,\mathrm{r},\uparrow} &= \biggr(\frac{\hbar v_x}{V_0}\biggr)n\pi, \label{eq:upres}\\
    W_{n,\mathrm{a},\uparrow} &=  \biggr(\frac{\hbar v_x}{V_0}\biggr)\frac{2n+1}{2}\pi, \label{eq:upares}
\end{align}
where it should be noted that a solution for a given $n$ is formally valid only if $n v_x/v_y\notin \mathbb{Z}$. In a similar fashion, we find that all solutions to Eq.~\eqref{Equation Resonance Condition 2} are of the form
\begin{align}
    W_{n,\mathrm{r},\downarrow} &= \biggr(\frac{\hbar v_y}{V_0}\biggr)n\pi,\label{Equation Down Resonance}\\
    W_{n,\mathrm{a},\downarrow} &=  \biggr(\frac{\hbar v_y}{V_0}\biggr)\frac{2n+1}{2}\pi,\label{Equation Down Anti-Resonance}
\end{align}
where a solution for a given $n$ is now valid only if $n v_y/v_x\notin \mathbb{Z}$. The subscript $r,\sigma$ is used to denote the resonant barrier widths that maximize the absolute value of the spin-$\sigma$ current, while the subscript $a,\sigma$ refers to the anti-resonant barrier widths that minimize said absolute value. The largest values of $|\mathcal{P}|$ are then obtained by choosing $W$ such that it is as close as possible to a resonant value for one spin-species, while simultaneously approaching an anti-resonant value for the other spin-species.

\begin{figure}
    \centering
    \includegraphics[width=\linewidth]{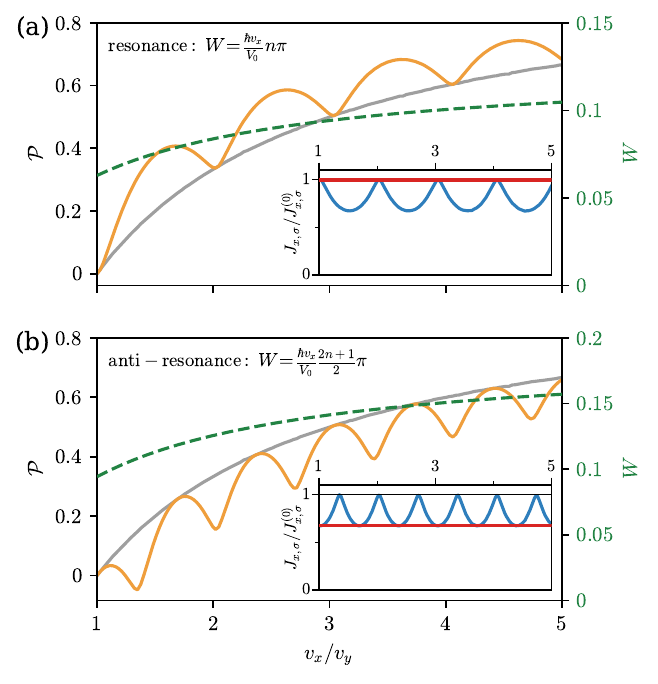}
    \caption{\textbf{Spin-current polarization for (anti-)resonant barrier widths.} (a) Polarization when spin-up tunneling is always resonant (yellow), compared to the polarization when the barrier is absent (grey). The width of the barrier changes as a function of $v_x$ according to Eq.~\eqref{eq:upres} with $n=1$, as depicted by the green line. We note that the ratio $v_x/v_y = (v_F+\xi)/(v_F-\xi)$ is varied by increasing $\xi$. As $v_x$ and $v_y$ are separately changing, $W$ does not follow a linear trend. The inset shows the spin-resolved current ratio between the barrier ($J_{x,\sigma}$) and no-barrier ($J^{(0)}_{x,\sigma}$) cases, with red for spin-up and blue for spin-down electrons. (b) Same as (a) but for spin-up tunneling always being anti-resonant, i.e. the width of the barrier is calculated from Eq.~\eqref{eq:upares}, with $n=1$. In both cases, $V_0=50$ (arb. units), $eV=0.02\,V_0$,  $v_F=1$ (arb. units) and $T=0$. }
    \label{fig:resonance}
\end{figure}

In Figs.~\ref{fig:resonance}(a) and (b), we plot the spin-current polarization as a function of the velocity anisotropy ratio for the $n=1$ resonance, Eq.~\eqref{eq:upres}, and anti-resonance widths, Eq.~\eqref{eq:upares}. For the resonance width, we observe a strong amplification of $\mathcal{P}$ compared to the situation without a potential barrier. For small values of $v_x/v_y$ in particular, the presence of the barrier can lead to increases of nearly $300\%$ in the spin-current polarization. Similarly, for the anti-resonance width, we find that the value of $\mathcal{P}$ is always lower compared to the situation without a potential barrier, and we even find that their signs can be opposite for small ratios $v_x/v_y$. Analogous results are obtained when considering the resonance and anti-resonance widths of Eqs.~\eqref{Equation Down Resonance} and \eqref{Equation Down Anti-Resonance}.


\section{Conclusion}\label{Section Conclusion}

We have shown that Klein tunneling in Dirac altermagnets can serve as an effective means to control and enhance spin-current polarization. For the $d$-wave Dirac altermagnet, we find that enhancements of up to a factor of 3 are readily achievable by a suitable tuning of the potential barrier parameters. For the $g$-wave Dirac altermagnet even more promising results are obtained, as it is seen that the presence of a potential barrier can significantly boost the spin-current polarization, even when the intrinsic polarization due to the spin-split band structure is vanishingly small. For a barrier implemented via electrostatic gating, such a mechanism would in turn allow the spin-current polarization to be switched on and off via a gate voltage. These results thus establish the potential of altermagnetic Klein tunneling for applications as a spin-current switch and amplifier.

Previous theoretical work has shown that Dirac points and altermagnetism can coexist \cite{Parshukov2025,LiuLizhou2025}, suggesting that materials behaving as $\ell$-wave Dirac altermagnets are physically realizable. Building on this, first-principles calculations in Ref.~\cite{Xu2025} identified several 2D candidate materials, including V$_2$STeO and Zr$_2$Br$_2$S, that exhibit both properties. Alternatively, one could also explore either 3D materials, as these may host altermagnetic Dirac surface states, or synthetic altermagnets \cite{Asgharpour2025}, for which a Dirac-like spectrum might be obtained via twist engineering.

Although we have made use of a specific minimal model to describe a Dirac altermagnet, we stress that the key results of this paper, namely the Klein-tunneling-induced spin-dependent transmission and altered spin-current polarization, are expected to be general characteristics of Dirac altermagnets. This is because these phenomena arise fundamentally from the interplay between a Dirac-like spectrum and $\ell$-wave symmetric spin splitting, which constitute the two defining features of these systems.

Beyond experimental realization, several interesting theoretical questions regarding this system remain open. For instance, while we have focused on electron transport, a similar framework could be applied to study magnon dynamics in Dirac altermagnets. Additionally, the impact of Klein tunneling on other magnetoelectric responses, such as the spin galvanic effect, has yet to be explored. A comprehensive study of these effects will be essential for identifying the distinct transport signatures emerging in Dirac altermagnets, potentially enabling highly efficient spin filtering and switching in next-generation spintronic devices.

Finally, our work highlights Klein tunneling in Dirac altermagnets as a promising mechanism for manipulating spin transport in quantum materials. The strong tunability of the spin-current polarization through barrier parameters and symmetry considerations suggests new avenues for engineering controllable spintronic functionality in systems where Dirac fermions and altermagnetic order coexist.

\section*{Acknowledgements}
The work of T.T.O. and R.A.D. was supported by the Dutch Research Council (NWO) by the research programme OCENW.XL21.XL21.058.
L.E. and C.M.S. acknowledge the research program Materials for the Quantum Age (QuMat) for financial support. This program (registration number 024.005.006) is part of the Gravitation program financed by the Dutch Ministry of Education, Culture and Science (OCW).

\bibliography{biblio}

\newpage
\appendix
\onecolumngrid

\section{Calculation of the transmission amplitude for a single elliptic Dirac Hamiltonian}\label{Appendix Transmission Amplitude}
Imposing continuity on the scattering wave solution, Eq.~\eqref{Equation Scattering Wave Solution}, at the boundaries $x=0$ and $x=W$, we arrive at the following set of linear equations,
\begin{align}
\begin{cases}
    1+r = \alpha + \beta,\\
    \alpha \, e^{i q_x W} + \beta \,e^{i q_r W} = t \, e^{i k_x W},\\
    s(e^{i\phi(\mathbf{k})}+r\,e^{i \phi(\mathbf{k}_r)}) = s' (\alpha \, e^{i \phi(\mathbf{q})}+\beta\,e^{i\phi(\mathbf{q}_r)}),\\
    s'(\alpha \, e^{i q_x W+i\phi(\mathbf{q})} + \beta \,e^{i q_r W+i\phi(\mathbf{q}_r)}) = s t \, e^{i k_x W+i\phi(\mathbf{k})}.
    \end{cases}
\end{align}
Using the second and the fourth equation of this set, we solve for $\alpha$ and $\beta$ in terms of $t$. We have
\begin{align}
    \begin{pmatrix}
        e^{i q_x W} &  e^{i q_r W}\\
         e^{i q_x W+i\phi(\mathbf{q})} & e^{i q_r W+i\phi(\mathbf{q}_r)}
    \end{pmatrix}
    \begin{pmatrix}
        \alpha\\
        \beta
    \end{pmatrix}
    &=
    t   e^{i k_x W}
    \begin{pmatrix}
        1\\
        ss'e^{i\phi(\mathbf{k})}
    \end{pmatrix}
    .
\end{align}
This $2\times 2$ system can be solved by means of Cramer's rule, which yields the following solution,
\begin{align}
    \begin{pmatrix}
        \alpha\\
        \beta
    \end{pmatrix}
    &=
    t\,e^{ik_xW}
    \begin{pmatrix}
        e^{-iq_xW}\bigg(\dfrac{1-ss'e^{i(\phi(\mathbf{k})-\phi(\mathbf{q}_r))}}{1-e^{i(\phi(\mathbf{q})-\phi(\mathbf{q}_r))}}\bigg)\\
        e^{-i q_r W}\bigg(\dfrac{1-ss'e^{i(\phi(\mathbf{k})-\phi(\mathbf{q}))}}{1-e^{i(\phi(\mathbf{q}_r)-\phi(\mathbf{q}))}}\bigg) 
    \end{pmatrix}
\end{align}
Plugging the solutions for $\alpha$ and $\beta$ back into the first and third equation of our set, we obtain
\begin{align}
    \begin{pmatrix}
        -1 & e^{i k_x W} \bigg[e^{-iq_xW}\bigg(\dfrac{1-ss'e^{i(\phi(\mathbf{k})-\phi(\mathbf{q}_r))}}{1-e^{i(\phi(\mathbf{q})-\phi(\mathbf{q}_r))}}\bigg)+ e^{-i q_r W}\bigg(\dfrac{1-ss'e^{i(\phi(\mathbf{k})-\phi(\mathbf{q}))}}{1-e^{i(\phi(\mathbf{q}_r)-\phi(\mathbf{q}))}}\bigg) \bigg]\\
        -e^{i\phi(\mathbf{k}_r)} & ss'e^{i k_x W} \bigg[e^{-iq_xW+i\phi(\mathbf{q})}\bigg(\dfrac{1-ss'e^{i(\phi(\mathbf{k})-\phi(\mathbf{q}_r))}}{1-e^{i(\phi(\mathbf{q})-\phi(\mathbf{q}_r))}}\bigg)+ e^{-i q_r W+i\phi(\mathbf{q}_r)}\bigg(\dfrac{1-ss'e^{i(\phi(\mathbf{k})-\phi(\mathbf{q}))}}{1-e^{i(\phi(\mathbf{q}_r)-\phi(\mathbf{q}))}} \bigg)\bigg]
    \end{pmatrix}
    \begin{pmatrix}
        r\\
        t
    \end{pmatrix}
    &=
    \begin{pmatrix}
        1\\
        e^{i\phi(\mathbf{k})}
    \end{pmatrix}
    .
\end{align}
We directly solve for $t$ by means of Cramer's rule, from which we find
\begin{align}\label{Equation Transmission Amplitude Exact}
    t &= \frac{\det
    \begin{pmatrix}
        -1 & 1 \\
        -e^{i\phi(\mathbf{k}_r)} & e^{i \phi(\mathbf{k})}
    \end{pmatrix}
    }{\det
    \begin{pmatrix}
        -1 & e^{i k_x W} \bigg[e^{-iq_xW}\bigg(\dfrac{1-ss'e^{i(\phi(\mathbf{k})-\phi(\mathbf{q}_r))}}{1-e^{i(\phi(\mathbf{q})-\phi(\mathbf{q}_r))}}\bigg)+ e^{-i q_r W}\bigg(\dfrac{1-ss'e^{i(\phi(\mathbf{k})-\phi(\mathbf{q}))}}{1-e^{i(\phi(\mathbf{q}_r)-\phi(\mathbf{q}))}}\bigg) \bigg]\\
        -e^{i\phi(\mathbf{k}_r)} & ss'e^{i k_x W} \bigg[e^{-iq_xW+i\phi(\mathbf{q})}\bigg(\dfrac{1-ss'e^{i(\phi(\mathbf{k})-\phi(\mathbf{q}_r))}}{1-e^{i(\phi(\mathbf{q})-\phi(\mathbf{q}_r))}}\bigg)+ e^{-i q_r W+i\phi(\mathbf{q}_r)}\bigg(\dfrac{1-ss'e^{i(\phi(\mathbf{k})-\phi(\mathbf{q}))}}{1-e^{i(\phi(\mathbf{q}_r)-\phi(\mathbf{q}))}} \bigg)\bigg]
    \end{pmatrix}
    }.
\end{align}
One can in principle proceed to work out the $2\times 2$ determinants given above, but it is obvious that in general the formula for the transmission amplitude $t$ will not be of a simple form. However, in the Klein tunneling limit where $|V_0| \gg |E|$, a relatively simple formula for $t$ can be obtained, and we therefore focus on this special case. Defining
\begin{align}
    \chi =\arg\bigg\{ v_x \cos(\theta) + i v_y \sin(\theta) \bigg\} ,
\end{align}
we find that in said limit we have
\begin{align}
    \phi(\mathbf{q}) &\approx \chi,\\
    \phi(\mathbf{q}_r) &\approx \chi + \pi.
\end{align}
Consequently, we find
\begin{align*}
    t &\approx \frac{e^{i \phi(\mathbf{k}_r)} -e^{i \phi(\mathbf{k})}
    }{\det
    \begin{pmatrix}
        -1 & e^{i (k_x-\eta) W} \bigg[e^{-iqW}\bigg(\dfrac{1+ss'e^{i(\phi(\mathbf{k})-\chi)}}{2}\bigg)+ e^{i q W}\bigg(\dfrac{1-ss'e^{i(\phi(\mathbf{k})-\chi)}}{2}\bigg) \bigg]\\
        -e^{i\phi(\mathbf{k}_r)} & ss'e^{i (k_x-\eta) W} \bigg[e^{-iq W+i\chi}\bigg(\dfrac{1+ss'e^{i(\phi(\mathbf{k})-\chi)}}{2}\bigg)- e^{i q W+i\chi}\bigg(\dfrac{1-ss'e^{i(\phi(\mathbf{k})-\chi)}}{2} \bigg)\bigg]
    \end{pmatrix}
    }.
\end{align*}
We can directly simplify the expressions inside the determinant, finding
\begin{align*}
    t &\approx \frac{\bigg[e^{i \phi(\mathbf{k}_r)} -e^{i \phi(\mathbf{k})} \bigg] \, e^{- i (k_x-\eta)W}
    }{\det
    \begin{pmatrix}
        -1 & \cos(q W) - i s s' e^{i (\phi(\mathbf{k})-\chi)} \sin(q W) \\
        -e^{i\phi(\mathbf{k}_r)} & ss' e^{i \chi} \bigg[-i\sin(q W) + s s' e^{i (\phi(\mathbf{k})-\chi)} \cos(q W) \bigg]
    \end{pmatrix}
    }.
\end{align*}
Working out this determinant, we then arrive at the final expression
\begin{align}
    t &\approx \frac{\bigg[e^{i \phi(\mathbf{k}_r)} -e^{i \phi(\mathbf{k})} \bigg]e^{-i (k_x-\eta) W}}{\cos(q W) \bigg[e^{i \phi(\mathbf{k}_r)} -e^{i \phi(\mathbf{k})} \bigg] + i s s' \sin(q W) \bigg[e^{i \chi} -e^{-i \chi}\, e^{i(\phi(\mathbf{k}_r) + \phi(\mathbf{k}))} \bigg]}.
\end{align}
The transmission coefficient is given by $\mathcal{T} = |t|^2$.

\section{Implementation of the $g$-wave Dirac altermagnet}\label{Appendix G-Wave}
For the $g$-wave Dirac altermagnet, we consider the spin-up Hamiltonian
\begin{equation}
    \hat{\mathcal{H}} = v_F \hat{p}_x \tau_x + v_F \hat{p}_y \tau_y + \xi (\hat{p}_x^2 - \hat{p}_y^2)^2\tau_z,
\end{equation}
where we note that the corresponding spin-down Hamiltonian is related to the above result by a rotation of $45^{\circ}$.
Following Eq.~\eqref{eq:rotp}, we can generalize to the case of an arbitrary angle $\theta$ w.r.t. the normal of the potential barrier, thus finding
\begin{equation}
    \hat{\mathcal{H}} = v_F (\hat{p}_x \cos \theta - \hat{p}_y \sin\theta) \tau_x + v_F (\hat{p}_y \cos \theta + \hat{p}_x \sin \theta) \tau_y + \xi \left[ (\hat{p}_x \cos \theta - \hat{p}_y \sin\theta)^2 - (\hat{p}_y \cos \theta + \hat{p}_x \sin \theta)^2 \right]^2\tau_z.
\end{equation}
The corresponding dispersion relation is readily obtained by diagonalizing,
\begin{equation}
    E(\mathbf{k}) = \pm \hbar \sqrt{v_F^2(k_x \cos \theta - k_y \sin\theta)^2  + v_F^2 (k_y \cos \theta + k_x \sin \theta)^2+ \hbar^2\xi^2 \left[ (k_x \cos \theta - k_y \sin\theta)^2 - (k_y \cos \theta + k_x \sin \theta)^2 \right]^4},
\end{equation}
with the corresponding group velocity components then given by
\begin{align*}
    v^g_{x}(\mathbf{k}) &= \frac{\hbar}{E(\mathbf{k})} \left\{v_F^2\cos\theta(k_x \cos \theta - k_y \sin\theta) + v_F^2 \sin \theta (k_y \cos \theta + k_x \sin \theta) \right.\\
     &+ \left. 2 \hbar^2\xi^2 \left[ (k_x \cos \theta - k_y \sin\theta)^2 - (k_y \cos \theta + k_x \sin \theta)^2 \right]^3 \left[      2 \cos \theta (k_x \cos \theta - k_y \sin\theta) - 2\sin\theta(k_y \cos \theta + k_x \sin \theta)  \right] \right\},
\end{align*}
and
\begin{align*}
    v^g_{y}(\mathbf{k}) &= \frac{\hbar}{E(\mathbf{k})} \left\{-v_F^2\sin\theta(k_x \cos \theta - k_y \sin\theta) + v_F^2 \cos \theta (k_y \cos \theta + k_x \sin \theta) \right.\\
     &+ \left. 2\hbar^2\xi^2 \left[ (k_x \cos \theta - k_y \sin\theta)^2 - (k_y \cos \theta + k_x \sin \theta)^2 \right]^3 \left[      -2 \sin \theta (k_x \cos \theta - k_y \sin\theta) - 2\cos\theta(k_y \cos \theta + k_x \sin \theta)  \right] \right\}.
\end{align*}

For the Schrödinger equation, including potential, i.e. $\hat{\mathcal{H}} \boldsymbol{\Psi} = [E-V(x)] \boldsymbol{\Psi}$, we now have
\begin{equation}
    \left[\sum_{n=0}^4 C_n \hat{p}_x^n \right] \boldsymbol{\Psi} = \mathbf{0} \quad \Leftrightarrow \quad \left[\sum_{n=0}^4 C_n (-i\hbar\partial_x)^n \right] \boldsymbol{\Psi} = \mathbf{0}, \label{eq:ODEG}
\end{equation}
where the coefficients $C_n$ are given by
\begin{align*}
    C_0 &= [V(x)-E] \mathbb{I} - \hbar v_F  k_y \sin\theta\, \tau_x + \hbar v_F k_y\,\cos\theta\, \tau_y + \frac{\xi}{2} \left[ \cos(4\theta) +1\right]\,\hbar^4  k_y^4 \,\tau_z,\\
    C_1 &= v_F \cos\theta\, \tau_x + v_F \sin\theta \, \tau_y + 2 \xi \sin(4\theta)\,\hbar^3 k_y^3\, \tau_z,\\
    C_2 &= - \xi \left[ 3 \cos(4\theta) -1 \right] \hbar^2k_y^2 \tau_z,\\
    C_3 &= -2\sin(4\theta)\,\hbar k_y \, \tau_z,\\
    C_4 &= \frac{\xi}{2} \left[ \cos(4\theta) +1\right] \tau_z.
\end{align*}
Equation ~\eqref{eq:ODEG} constitutes a fourth-order ODE, and we therefore need four boundary conditions at the barrier interfaces. We will require the wavefunction, and its first three derivatives, to be continuous across the interface, i.e.
\begin{align}
    \lim_{x\to x_0^-} \boldsymbol{\Psi}(x,y) &= \lim_{x\to x_0^+} \boldsymbol{\Psi}(x,y),\notag\\
    \lim_{x\to x_0^-} \boldsymbol{\Psi}'(x,y) &= \lim_{x\to x_0^+} \boldsymbol{\Psi}'(x,y),\notag\\
    \lim_{x\to x_0^-} \boldsymbol{\Psi}''(x,y) &= \lim_{x\to x_0^+} \boldsymbol{\Psi}''(x,y), \notag\\
    \lim_{x\to x_0^-} \boldsymbol{\Psi}'''(x,y) &= \lim_{x\to x_0^+} \boldsymbol{\Psi}'''(x,y), \label{eq:BC}
\end{align}
for $x_0 \in \{0,W \}$. Upon defining $\boldsymbol{\Phi}\equiv (\boldsymbol{\Psi},\boldsymbol{\Psi}',\boldsymbol{\Psi}'',\boldsymbol{\Psi}''')^T$, we cast eq.~\eqref{eq:ODEG} in the following form
\begin{equation}
    \begin{pmatrix}
        \mathbb{I} & 0 & 0 & 0 \\
        0 & \mathbb{I} & 0 & 0 \\
        0 & 0 & \mathbb{I} & 0 \\
        0 & 0 & 0 & C_4 
    \end{pmatrix}
    \boldsymbol{\Phi}' = 
    \begin{pmatrix}
        0 & \mathbb{I} & 0 & 0 \\
        0 & 0 & \mathbb{I} & 0 \\
        0 & 0 & 0 & \mathbb{I} \\
        -C_0 & iC_1  & C_2  & -i C_3
    \end{pmatrix}\boldsymbol{\Phi} \quad \leftrightarrow \quad A\boldsymbol{\Phi}' = M \boldsymbol{\Phi}, \label{eq:matrixODE}
\end{equation}
with $\mathbb{I}$ the $2\times 2$ identity matrix. For the special angles $\theta = (2m+1)\pi/4$, with $m \in \mathbb{Z}$, we have $C_4 =C_3=0$, in which case the Hamiltonian is only second order in $k_x$;
\begin{equation}
    \begin{pmatrix}
        \mathbb{I} & 0 \\
        0 & -C_2 
    \end{pmatrix}
    \begin{pmatrix}
        \boldsymbol{\Psi}' \\ \boldsymbol{\Psi}''
    \end{pmatrix}
    =
    \begin{pmatrix}
        0 & \mathbb{I} \\
         -C_0 & iC_1
    \end{pmatrix}
    \begin{pmatrix}
        \boldsymbol{\Psi} \\ \boldsymbol{\Psi}'
    \end{pmatrix}.
\end{equation}
Now, Eq.~\eqref{eq:matrixODE} admits solutions of the form 
\begin{equation}
    \boldsymbol{\Phi} = \sum_{i=1}^8 \alpha_i e^{\lambda_i x} \boldsymbol{\chi}_i,
\end{equation}
where the coefficients $\alpha_i$ are to be determined from the boundary conditions and $\lambda_i, \boldsymbol{\chi}_i$ satisfy $A^{-1}M \boldsymbol{\chi}_i = \lambda_i \boldsymbol{\chi}_i$. A solution for which $\text{Re}(\lambda_i)=0$ is propagating while $\text{Re}(\lambda_i)\neq0$ corresponds to an evanescent wave. For sufficiently small $\xi$, there is only one incoming/outgoing propagating mode, such that we write
\begin{equation}
    \boldsymbol{\Phi}_\text{I} = e^{ik_\text{in}x} \boldsymbol{\chi}_\text{in} + r e^{ik_\text{ref}x} \boldsymbol{\chi}_\text{ref} + \sum_{\text{Re}\left(\lambda_j^{(\text{I})}\right)>0} \alpha_j^{(\text{I})} e^{  \lambda_j^{(I)} x} \boldsymbol{\chi}_j^{(\text{I})},
\end{equation}
where we only take evanescent modes with $\text{Re}(\lambda_j)>0$ because those with $\text{Re}(\lambda_j)<0$ are not normalizable. In region II, we consider all modes
\begin{equation}
\boldsymbol{\Phi}_{\text{II}} = \sum_{j=1}^8 \alpha_j^{({\text{II}})} e^{\lambda_j^{({\text{II}})} x} \boldsymbol{\chi}_j^{({\text{II}})}.
\end{equation}
Finally, in region III, we consider one transmitted mode and the evanescent modes that are normalizable, i.e. $\text{Re} \left( \lambda_i^{(\text{III})} \right) < 0$;
\begin{equation}
    \boldsymbol{\Phi}_\text{III} = te^{ik_\text{tr}x} \boldsymbol{\chi}_\text{tr} +\sum_{\text{Re}\left(\lambda_j^{(\text{III})}\right)<0} \alpha_j^{(\text{III})} e^{  \lambda_j^{(\text{III})} x} \boldsymbol{\chi}^{(\text{III})}_j.
\end{equation}
These equations, subject to the boundary conditions in eq.~\eqref{eq:BC}, are solved numerically to obtain the transmission probability $|t|^2$. The results for the spin-down transmission are readily obtained by taking $\theta \to \theta+45^\circ$.

\end{document}